%% file: main.tex
\documentclass[sigconf,natbib=true]{acmart}

\AtBeginDocument{%
  }

\copyrightyear{2023}
\acmYear{2023}
\setcopyright{rightsretained}
\acmConference[SIGIR '23]{Proceedings of the 46th International ACM SIGIR Conference on Research and Development in Information Retrieval}{July 23--27, 2023}{Taipei, Taiwan}
\acmBooktitle{Proceedings of the 46th International ACM SIGIR Conference on Research and Development in Information Retrieval (SIGIR '23), July 23--27, 2023, Taipei, Taiwan}\acmDOI{10.1145/3539618.3591881}
\acmISBN{978-1-4503-9408-6/23/07}

\usepackage[utf8]{inputenc} 
\usepackage[T1]{fontenc}    
\usepackage{hyperref}       
\usepackage{url}            
\usepackage{booktabs}       
\usepackage{amsmath}        
\usepackage{amsfonts}       
\usepackage{nicefrac}       
\usepackage{microtype}      
\usepackage{xcolor}         
\usepackage{enumitem} 
\usepackage{float}
\usepackage{graphicx}
\usepackage{caption}
\usepackage{subcaption} 
\usepackage{tabularx}
\usepackage{multirow}
\usepackage{algorithm}
\usepackage{algpseudocode}
\usepackage{soul}
\usepackage{tikz} 
\usepackage{pgfplots}
\usepackage{pifont}

\newcommand{\todo}[1]{\textcolor{red}{#1}}

\begin{document}

\title{
Beyond Single Items: Exploring User Preferences in Item Sets with the Conversational Playlist Curation Dataset
}

\author{\mbox{Arun Tejasvi Chaganty}}
\affiliation{%
  \institution{Google Research}
  \country{}
 }
\email{arunchaganty@google.com}

\author{Megan Leszczynski}
\authornote{Work done while interning at Google Research.}
\affiliation{%
  \institution{Stanford University}
    \country{} 
}
\email{mleszczy@cs.stanford.edu}

\author{Shu Zhang}
\affiliation{%
  \institution{Google Research}
    \country{}
}
\email{shzhang@google.com}

\author{Ravi Ganti}
\affiliation{%
  \institution{Google Research}
    \country{}
}
\email{gmravi@google.com}

\author{Krisztian Balog}
\affiliation{%
 \institution{Google Research}
   \country{}
 }
 \email{krisztianb@google.com}

\author{Filip Radlinski}
\affiliation{%
 \institution{Google Research}
   \country{}
 }
\email{filiprad@google.com}

\renewcommand{\shortauthors}{Arun Tejasvi Chaganty et al.}




\input{std-macros}
\input{macros}

\input{00_abstract}

\maketitle

\input{01_intro}
\input{02_related}
\input{03_task}
\input{04_protocol}
\input{05_analysis}
\input{06_baselines}
\input{07_discussion}

\bibliographystyle{ACM-Reference-Format}
\bibliography{custom}

\appendix

\end{document}

%% file: std-macros.tex
\providecommand\sa{\ensuremath{\mathcal{a}}}
\providecommand\sd{\ensuremath{\mathcal{d}}}
\providecommand\se{\ensuremath{\mathcal{e}}}
\providecommand\sg{\ensuremath{\mathcal{g}}}
\providecommand\sh{\ensuremath{\mathcal{h}}}
\providecommand\si{\ensuremath{\mathcal{i}}}
\providecommand\sj{\ensuremath{\mathcal{j}}}
\providecommand\sk{\ensuremath{\mathcal{k}}}
\providecommand\sm{\ensuremath{\mathcal{m}}}
\providecommand\sn{\ensuremath{\mathcal{n}}}
\providecommand\so{\ensuremath{\mathcal{o}}}
\providecommand\sq{\ensuremath{\mathcal{q}}}
\providecommand\sr{\ensuremath{\mathcal{r}}}
\providecommand\st{\ensuremath{\mathcal{t}}}
\providecommand\su{\ensuremath{\mathcal{u}}}
\providecommand\sv{\ensuremath{\mathcal{v}}}
\providecommand\sw{\ensuremath{\mathcal{w}}}
\providecommand\sx{\ensuremath{\mathcal{x}}}
\providecommand\sy{\ensuremath{\mathcal{y}}}
\providecommand\sz{\ensuremath{\mathcal{z}}}
\providecommand\sA{\ensuremath{\mathcal{A}}}
\providecommand\sB{\ensuremath{\mathcal{B}}}
\providecommand\sC{\ensuremath{\mathcal{C}}}
\providecommand\sD{\ensuremath{\mathcal{D}}}
\providecommand\sE{\ensuremath{\mathcal{E}}}
\providecommand\sF{\ensuremath{\mathcal{F}}}
\providecommand\sG{\ensuremath{\mathcal{G}}}
\providecommand\sH{\ensuremath{\mathcal{H}}}
\providecommand\sI{\ensuremath{\mathcal{I}}}
\providecommand\sJ{\ensuremath{\mathcal{J}}}
\providecommand\sK{\ensuremath{\mathcal{K}}}
\providecommand\sL{\ensuremath{\mathcal{L}}}
\providecommand\sM{\ensuremath{\mathcal{M}}}
\providecommand\sN{\ensuremath{\mathcal{N}}}
\providecommand\sO{\ensuremath{\mathcal{O}}}
\providecommand\sP{\ensuremath{\mathcal{P}}}
\providecommand\sQ{\ensuremath{\mathcal{Q}}}
\providecommand\sR{\ensuremath{\mathcal{R}}}
\providecommand\sS{\ensuremath{\mathcal{S}}}
\providecommand\sT{\ensuremath{\mathcal{T}}}
\providecommand\sU{\ensuremath{\mathcal{U}}}
\providecommand\sV{\ensuremath{\mathcal{V}}}
\providecommand\sW{\ensuremath{\mathcal{W}}}
\providecommand\sX{\ensuremath{\mathcal{X}}}
\providecommand\sY{\ensuremath{\mathcal{Y}}}
\providecommand\sZ{\ensuremath{\mathcal{Z}}}
\providecommand\ba{\ensuremath{\mathbf{a}}}
\providecommand\bb{\ensuremath{\mathbf{b}}}
\providecommand\bc{\ensuremath{\mathbf{c}}}
\providecommand\bd{\ensuremath{\mathbf{d}}}
\providecommand\be{\ensuremath{\mathbf{e}}}
\providecommand\bg{\ensuremath{\mathbf{g}}}
\providecommand\bh{\ensuremath{\mathbf{h}}}
\providecommand\bi{\ensuremath{\mathbf{i}}}
\providecommand\bj{\ensuremath{\mathbf{j}}}
\providecommand\bk{\ensuremath{\mathbf{k}}}
\providecommand\bl{\ensuremath{\mathbf{l}}}
\providecommand\bn{\ensuremath{\mathbf{n}}}
\providecommand\bo{\ensuremath{\mathbf{o}}}
\providecommand\bp{\ensuremath{\mathbf{p}}}
\providecommand\bq{\ensuremath{\mathbf{q}}}
\providecommand\br{\ensuremath{\mathbf{r}}}
\providecommand\bs{\ensuremath{\mathbf{s}}}
\providecommand\bt{\ensuremath{\mathbf{t}}}
\providecommand\bu{\ensuremath{\mathbf{u}}}
\providecommand\bv{\ensuremath{\mathbf{v}}}
\providecommand\bw{\ensuremath{\mathbf{w}}}
\providecommand\bx{\ensuremath{\mathbf{x}}}
\providecommand\by{\ensuremath{\mathbf{y}}}
\providecommand\bz{\ensuremath{\mathbf{z}}}
\providecommand\bA{\ensuremath{\mathbf{A}}}
\providecommand\bB{\ensuremath{\mathbf{B}}}
\providecommand\bC{\ensuremath{\mathbf{C}}}
\providecommand\bD{\ensuremath{\mathbf{D}}}
\providecommand\bE{\ensuremath{\mathbf{E}}}
\providecommand\bF{\ensuremath{\mathbf{F}}}
\providecommand\bG{\ensuremath{\mathbf{G}}}
\providecommand\bH{\ensuremath{\mathbf{H}}}
\providecommand\bI{\ensuremath{\mathbf{I}}}
\providecommand\bJ{\ensuremath{\mathbf{J}}}
\providecommand\bK{\ensuremath{\mathbf{K}}}
\providecommand\bL{\ensuremath{\mathbf{L}}}
\providecommand\bM{\ensuremath{\mathbf{M}}}
\providecommand\bN{\ensuremath{\mathbf{N}}}
\providecommand\bO{\ensuremath{\mathbf{O}}}
\providecommand\bP{\ensuremath{\mathbf{P}}}
\providecommand\bQ{\ensuremath{\mathbf{Q}}}
\providecommand\bR{\ensuremath{\mathbf{R}}}
\providecommand\bS{\ensuremath{\mathbf{S}}}
\providecommand\bT{\ensuremath{\mathbf{T}}}
\providecommand\bU{\ensuremath{\mathbf{U}}}
\providecommand\bV{\ensuremath{\mathbf{V}}}
\providecommand\bW{\ensuremath{\mathbf{W}}}
\providecommand\bX{\ensuremath{\mathbf{X}}}
\providecommand\bY{\ensuremath{\mathbf{Y}}}
\providecommand\bZ{\ensuremath{\mathbf{Z}}}
\providecommand\Ba{\ensuremath{\mathbb{a}}}
\providecommand\Bb{\ensuremath{\mathbb{b}}}
\providecommand\Bc{\ensuremath{\mathbb{c}}}
\providecommand\Bd{\ensuremath{\mathbb{d}}}
\providecommand\Be{\ensuremath{\mathbb{e}}}
\providecommand\Bf{\ensuremath{\mathbb{f}}}
\providecommand\Bg{\ensuremath{\mathbb{g}}}
\providecommand\Bh{\ensuremath{\mathbb{h}}}
\providecommand\Bi{\ensuremath{\mathbb{i}}}
\providecommand\Bj{\ensuremath{\mathbb{j}}}
\providecommand\Bk{\ensuremath{\mathbb{k}}}
\providecommand\Bl{\ensuremath{\mathbb{l}}}
\providecommand\Bm{\ensuremath{\mathbb{m}}}
\providecommand\Bn{\ensuremath{\mathbb{n}}}
\providecommand\Bo{\ensuremath{\mathbb{o}}}
\providecommand\Bp{\ensuremath{\mathbb{p}}}
\providecommand\Bq{\ensuremath{\mathbb{q}}}
\providecommand\Br{\ensuremath{\mathbb{r}}}
\providecommand\Bs{\ensuremath{\mathbb{s}}}
\providecommand\Bt{\ensuremath{\mathbb{t}}}
\providecommand\Bu{\ensuremath{\mathbb{u}}}
\providecommand\Bv{\ensuremath{\mathbb{v}}}
\providecommand\Bw{\ensuremath{\mathbb{w}}}
\providecommand\Bx{\ensuremath{\mathbb{x}}}
\providecommand\By{\ensuremath{\mathbb{y}}}
\providecommand\Bz{\ensuremath{\mathbb{z}}}
\providecommand\BA{\ensuremath{\mathbb{A}}}
\providecommand\BB{\ensuremath{\mathbb{B}}}
\providecommand\BC{\ensuremath{\mathbb{C}}}
\providecommand\BD{\ensuremath{\mathbb{D}}}
\providecommand\BE{\ensuremath{\mathbb{E}}}
\providecommand\BF{\ensuremath{\mathbb{F}}}
\providecommand\BG{\ensuremath{\mathbb{G}}}
\providecommand\BH{\ensuremath{\mathbb{H}}}
\providecommand\BI{\ensuremath{\mathbb{I}}}
\providecommand\BJ{\ensuremath{\mathbb{J}}}
\providecommand\BK{\ensuremath{\mathbb{K}}}
\providecommand\BL{\ensuremath{\mathbb{L}}}
\providecommand\BM{\ensuremath{\mathbb{M}}}
\providecommand\BN{\ensuremath{\mathbb{N}}}
\providecommand\BO{\ensuremath{\mathbb{O}}}
\providecommand\BP{\ensuremath{\mathbb{P}}}
\providecommand\BQ{\ensuremath{\mathbb{Q}}}
\providecommand\BR{\ensuremath{\mathbb{R}}}
\providecommand\BS{\ensuremath{\mathbb{S}}}
\providecommand\BT{\ensuremath{\mathbb{T}}}
\providecommand\BU{\ensuremath{\mathbb{U}}}
\providecommand\BV{\ensuremath{\mathbb{V}}}
\providecommand\BW{\ensuremath{\mathbb{W}}}
\providecommand\BX{\ensuremath{\mathbb{X}}}
\providecommand\BY{\ensuremath{\mathbb{Y}}}
\providecommand\BZ{\ensuremath{\mathbb{Z}}}
\providecommand\balpha{\ensuremath{\mbox{\boldmath$\alpha$}}}
\providecommand\bbeta{\ensuremath{\mbox{\boldmath$\beta$}}}
\providecommand\btheta{\ensuremath{\mbox{\boldmath$\theta$}}}
\providecommand\bphi{\ensuremath{\mbox{\boldmath$\phi$}}}
\providecommand\bpi{\ensuremath{\mbox{\boldmath$\pi$}}}
\providecommand\bpsi{\ensuremath{\mbox{\boldmath$\psi$}}}
\providecommand\bmu{\ensuremath{\mbox{\boldmath$\mu$}}}
\newcommand\fig[1]{\begin{center} \includegraphics{#1} \end{center}}
\newcommand\Fig[4]{\begin{figure}[ht] \begin{center} \includegraphics[scale=#2]{#1} \end{center} \caption{\label{fig:#3} #4} \end{figure}}
\newcommand\FigTop[4]{\begin{figure}[t] \begin{center} \includegraphics[scale=#2]{#1} \end{center} \caption{\label{fig:#3} #4} \end{figure}}
\newcommand\FigStar[4]{\begin{figure*}[ht] \begin{center} \includegraphics[scale=#2]{#1} \end{center} \caption{\label{fig:#3} #4} \end{figure*}}
\newcommand\FigRight[4]{\begin{wrapfigure}{r}{0.5\textwidth} \begin{center} \includegraphics[width=0.5\textwidth]{#1} \end{center} \caption{\label{fig:#3} #4} \end{wrapfigure}}
\newcommand\aside[1]{\quad\text{[#1]}}
\newcommand\interpret[1]{\llbracket #1 \rrbracket} 
\newcommand{\var}{\text{Var}} 
\newcommand{\cov}{\text{Cov}} 
\newcommand\p[1]{\ensuremath{\left( #1 \right)}} 
\newcommand\pa[1]{\ensuremath{\left\langle #1 \right\rangle}} 
\newcommand\pb[1]{\ensuremath{\left[ #1 \right]}} 
\newcommand\pc[1]{\ensuremath{\left\{ #1 \right\}}} 
\newcommand\eval[2]{\ensuremath{\left. #1 \right|_{#2}}} 
\newcommand\inv[1]{\ensuremath{\frac{1}{#1}}}
\newcommand\half{\ensuremath{\frac{1}{2}}}
\newcommand\R{\ensuremath{\mathbb{R}}} 
\newcommand\Z{\ensuremath{\mathbb{Z}}} 
\newcommand\inner[2]{\ensuremath{\left< #1, #2 \right>}} 
\newcommand\mat[2]{\ensuremath{\left(\begin{array}{#1}#2\end{array}\right)}} 
\newcommand\eqn[1]{\begin{align} #1 \end{align}} 
\newcommand\eqnl[2]{\begin{align} \label{eqn:#1} #2 \end{align}} 
\newcommand\eqdef{\ensuremath{\stackrel{\rm def}{=}}} 
\newcommand{\1}{\mathbb{I}} 
\newcommand{\bone}{\mathbf{1}} 
\newcommand{\bzero}{\mathbf{0}} 
\newcommand\refeqn[1]{(\ref{eqn:#1})}
\newcommand\refeqns[2]{(\ref{eqn:#1}) and (\ref{eqn:#2})}
\newcommand\refchp[1]{Chapter~\ref{chp:#1}}
\newcommand\refchap[1]{Chapter~\ref{chap:#1}}
\newcommand\refsec[1]{Section~\ref{sec:#1}}
\newcommand\refsecs[2]{Sections~\ref{sec:#1} and~\ref{sec:#2}}
\newcommand\reffig[1]{Figure~\ref{fig:#1}}
\newcommand\reffigs[2]{Figures~\ref{fig:#1} and~\ref{fig:#2}}
\newcommand\reffigss[3]{Figures~\ref{fig:#1},~\ref{fig:#2}, and~\ref{fig:#3}}
\newcommand\reffigsss[4]{Figures~\ref{fig:#1},~\ref{fig:#2},~\ref{fig:#3}, and~\ref{fig:#4}}
\newcommand\reftab[1]{Table~\ref{tab:#1}}
\newcommand\refapp[1]{Appendix~\ref{sec:#1}}
\newcommand\refthm[1]{Theorem~\ref{thm:#1}}
\newcommand\refthms[2]{Theorems~\ref{thm:#1} and~\ref{thm:#2}}
\newcommand\reflem[1]{Lemma~\ref{lem:#1}}
\newcommand\reflems[2]{Lemmas~\ref{lem:#1} and~\ref{lem:#2}}
\newcommand\refalg[1]{Algorithm~\ref{alg:#1}}
\newcommand\refalgs[2]{Algorithms~\ref{alg:#1} and~\ref{alg:#2}}
\newcommand\refex[1]{Example~\ref{ex:#1}}
\newcommand\refexs[2]{Examples~\ref{ex:#1} and~\ref{ex:#2}}
\newcommand\refprop[1]{Proposition~\ref{prop:#1}}
\newcommand\refdef[1]{Definition~\ref{def:#1}}
\newcommand\refcor[1]{Corollary~\ref{cor:#1}}
\newcommand\Chapter[2]{\chapter{#2}\label{chp:#1}}
\newcommand\Section[2]{\section{#2}\label{sec:#1}}
\newcommand\Subsection[2]{\subsection{#2}\label{sec:#1}}
\newcommand\Subsubsection[2]{\subsubsection{#2}\label{sec:#1}}
\ifthenelse{\isundefined{\definition}}{\newtheorem{definition}{Definition}}{}
\ifthenelse{\isundefined{\assumption}}{\newtheorem{assumption}{Assumption}}{}
\ifthenelse{\isundefined{\hypothesis}}{\newtheorem{hypothesis}{Hypothesis}}{}
\ifthenelse{\isundefined{\proposition}}{\newtheorem{proposition}{Proposition}}{}
\ifthenelse{\isundefined{\theorem}}{\newtheorem{theorem}{Theorem}}{}
\ifthenelse{\isundefined{\lemma}}{\newtheorem{lemma}{Lemma}}{}
\ifthenelse{\isundefined{\corollary}}{\newtheorem{corollary}{Corollary}}{}
\ifthenelse{\isundefined{\alg}}{\newtheorem{alg}{Algorithm}}{}
\ifthenelse{\isundefined{\example}}{\newtheorem{example}{Example}}{}
\newcommand\cv{\ensuremath{\to}} 
\newcommand\cvL{\ensuremath{\xrightarrow{\mathcal{L}}}} 
\newcommand\cvd{\ensuremath{\xrightarrow{d}}} 
\newcommand\cvP{\ensuremath{\xrightarrow{P}}} 
\newcommand\cvas{\ensuremath{\xrightarrow{a.s.}}} 
\newcommand\eqdistrib{\ensuremath{\stackrel{d}{=}}} 
\newcommand{\E}{\ensuremath{\mathbb{E}}} 
\newcommand\KL[2]{\ensuremath{\text{KL}\left( #1 \| #2 \right)}} 

%% file: macros.tex

\newcommand{\editmode}{}

\ifundef{\editmode}{
\newcommand{\ac}[1]{}
\newcommand{\ml}[1]{}
\newcommand{\rg}[1]{}
\newcommand{\fr}[1]{}
\newcommand{\kb}[1]{}
\newcommand{\fp}[1]{}
\newcommand{\TODO}[1]{}
\newcommand{\edit}[2]{#1}
}{%
\newcommand{\ac}[1]{{\color{olive}{\bf{AC:}} \emph{#1}}}
\newcommand{\ml}[1]{{\color{olive}{\bf{ML:}} \emph{#1}}}
\newcommand{\rg}[1]{{\color{olive}{\bf{RG:}} \emph{#1}}}
\newcommand{\fr}[1]{{\color{olive}{\bf{FR:}} \emph{#1}}}
\newcommand{\kb}[1]{{\color{olive}{\bf{KB:}} \emph{#1}}}
\newcommand{\fp}[1]{{\color{olive}{\bf{FP:}} \emph{#1}}}
\newcommand{\TODO}[1]{{\color{magenta}{\bf{TODO}} \emph{#1}}}
\newcommand{\edit}[2]{{\color{blue} #1}{\st{#2}}}
}

\newcommand{\Xstar}{X_{*}}
\newcommand{\bbR}{\mathbb{R}}

\newcommand{\ourdata}{CPCD}
\newcommand{\playlistdata}{ExpertPlaylists} 
\newcommand{\UserTemplates}{\texttt{UserTemplates}}

\newcommand{\ourmethod}{TalkTheWalk}
\newcommand{\shortourmethod}{TtW}
\newcommand{\sysname}{DE$\rhd$\ourdata{}}
\newcommand{\csysname}{Contriever$\rhd$\ourdata{}}
\newcommand{\shortsysname}{$\rhd$\ourdata{}}
\newcommand{\oursystem}{\sysname{}}
\newcommand{\shortoursystem}{\shortsysname{}}
\newcommand{\nonconvo}{DE$\rhd$\playlistdata{}}
\newcommand{\nonconvoDE}{item collection DE}

\newcommand{\mR}{R}

\newcommand{\xr}{\rho}
\newcommand{\xE}{P}

\newcommand{\sep}{\texttt{[SEP]}\ }

\newcommand{\xe}{\xr}

\newcommand{\xu}{u}

\newcommand{\xx}{x}
\newcommand{\xs}{\bs}
\newcommand{\xp}{\bz}
\newcommand{\xc}{\br}
\newcommand{\xxv}{\tilde{\xx}}
\newcommand{\xpv}{\tilde{\xp}}
\newcommand{\xsv}{\tilde{\xs}}
\newcommand{\xcv}{\tilde{\xc}}

\newcommand{\cp} {conversational retrieval~}

\newcommand{\embed}{\operatorname{\textrm{embed}}}
\newcommand*{\defeq}{\stackrel{\text{def}}{=}}


\newcommand{\revision}[2]{\textst{#1} {\color{blue} #2}}
\newcommand{\edited}[1]{{\color{red} #1}}

\newcommand{\desc}{\textbf{\textit{\$description}}}

\newcommand{\cmark}{\ding{51}}%
\newcommand{\xmark}{\ding{55}}%
\newcommand{\music}[1]{\href{https://music.youtube.com/watch?v=#1}{\textmusicalnote}}%

\newcommand{\dataseturl}{{\color{blue} \url{https://github.com/google-research-datasets/cpcd}}}

%% file: 00_abstract.tex
\begin{abstract}
Users in consumption domains, like music, are often able to more efficiently provide preferences over a set of items (e.g. a playlist or radio) than over single items (e.g. songs).
Unfortunately, this is an underexplored area of research, with most existing recommendation systems limited to understanding preferences over single items.
Curating an item set exponentiates the search space that recommender systems must consider (all subsets of items!):
this motivates conversational approaches---where users explicitly state or refine their preferences and systems elicit preferences in natural language---as an efficient way to understand user needs.
We call this task conversational item set curation and present a novel data collection methodology that efficiently collects realistic preferences about item sets in a conversational setting by observing both item-level and set-level feedback.
We apply this methodology to music recommendation to build the Conversational Playlist Curation Dataset (CPCD)\footnote{%
The dataset can be viewed at \dataseturl{}.
}, where we
show that it leads raters to express preferences that would not be otherwise expressed.
Finally, we propose a wide range of conversational retrieval models as baselines for this task and evaluate them on the dataset.
\end{abstract}


%% file: 01_intro.tex
\Section{intro}{Introduction}
\begin{figure}
    \centering
    \includegraphics[width=\columnwidth]{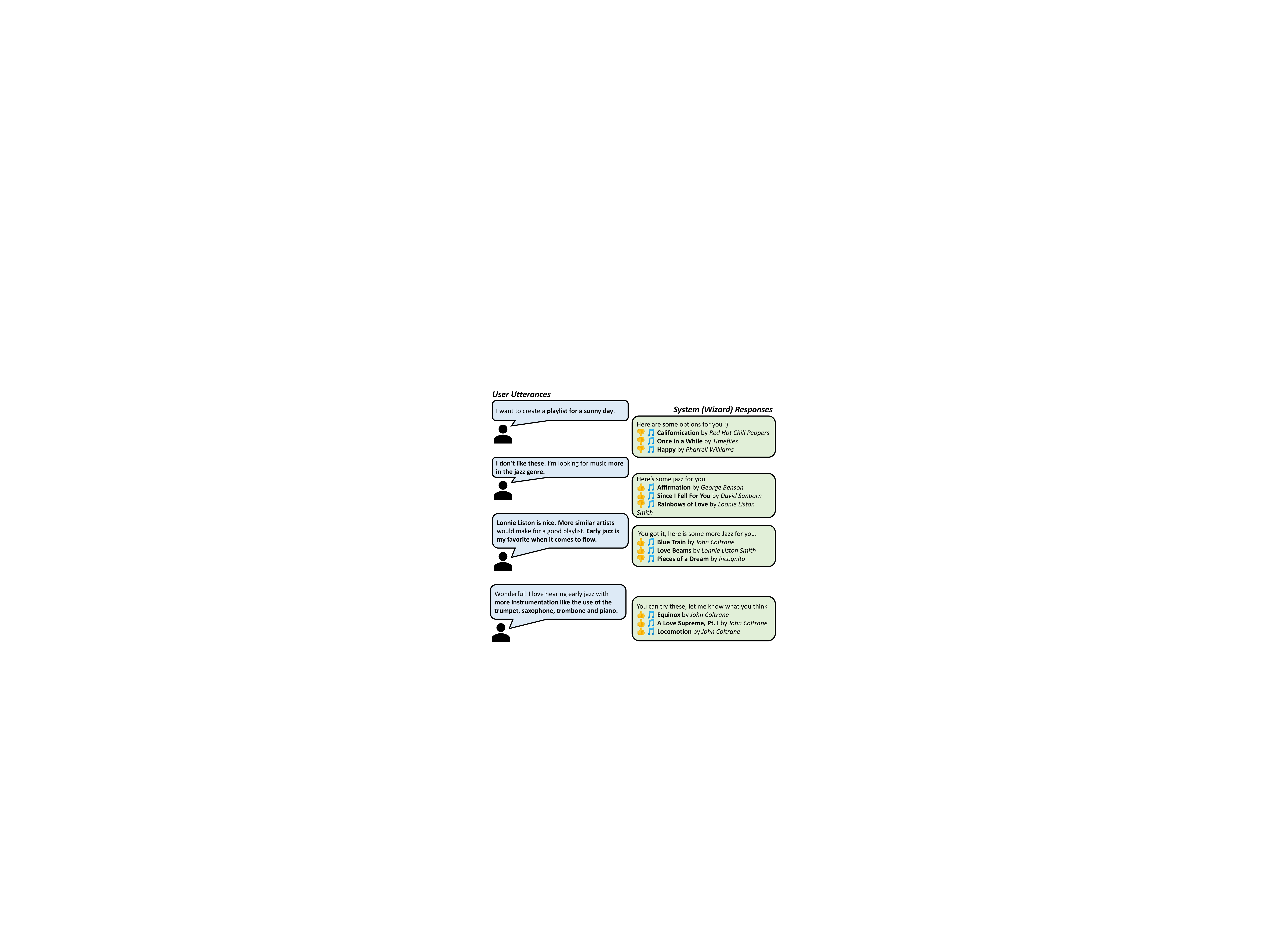}
    \caption{
    An example conversation from the dataset.
    In conversational item set curation, users collaborate with the system to iteratively curate an \textit{item set} as opposed to a single item.
    As a result, user preferences often apply at the set-level (e.g., ``a playlist for a sunny day'') and center on thematic cohesion (e.g., ``Early jazz is my favorite when it comes to flow'') and diversity (``More similar artists'').
    }
    \label{fig:example}
    \vspace{-1em}
\end{figure}

In recent years, conversational agents have become increasingly popular for a variety of tasks, including recommendation systems \cite{cis-survey,jannach-convrec-survey}. 
The conversational paradigm is especially suited for recommendation tasks, because of the convenience of using natural language to express preferences and provide feedback~\citep{zhang2018towards,Zou:2020:SIGIR,Xu:2021:WSDM,Kostric:2021:RecSys}.
Most existing conversational recommendation methods focus on recommending individual items~\citep{jannach-convrec-survey}.
However, there are also scenarios where the user aims to create a set of items from a larger collection---creating a music playlist for a given context, like working out or for a long drive, is a common manifestation of this problem.
In this setting, the goal of the conversational agent is to aid the user in the selection of a \emph{set of items} that the user has in mind.
The problem of recommending sets of items in a conversational setting, which we refer to as \emph{conversational item set curation}, is still largely unexplored.
We use the term curation to highlight that this is a collaborative and iterative process that requires the conversational agent to elicit and understand the user's preferences in order to effectively aid them.
In this work, we propose a framing of this novel problem, create a representative dataset, and present and experimentally evaluate a set of baselines.

The problem of conversational item set curation differs from prior work on entity set retrieval~\citep{Bron:2013:EBE,Zhang:2017:ESA,He:2011:WWW,Ciglan:2012:WWW,Zhang:2017:SIGIR} in a number of important ways.
First, what constitutes a relevant item for a set is highly subjective; different people would find different music ``energetic'' or ``relaxing'' for a given context.  The conversational setting allows for natural feedback both on the item and on the set level.  
Second, the curation of the item set is more iterative and interactive than entity set retrieval, as the conversational agent must engage in a back-and-forth dialogue with the user in order to fully understand their preferences. This requires the development of more sophisticated dialog management techniques, including effective strategies for balancing exploration and exploitation as well as the ability to handle uncertainty and ambiguity in the user's responses. 
We break down the problem of conversational item set curation into a set of more specific tasks, in order to provide a clear and specific focus for further research in this space.
Specifically, 
(1) \emph{conversational item retrieval} is the task of returning relevant items from a collection based on the conversation history;
(2) \emph{conversational item set refinement} identifies which and how many retrieved items to use to update the item set; and
(3) \emph{conversational elicitation/explanation generation} is concerned with the generation of questions for eliciting user preferences either via direct questions or indirectly via user feedback provided on the explanations that accompany the system's recommendations.

A prerequisite first step to allow research on the above tasks is the construction of a dataset with the desired characteristics.
We present a novel human-to-human data collection methodology that efficiently acquires realistic preferences about item sets that are incrementally constructed, observing both item-level and set-level feedback. 
Specifically, we perform the data collection within the domain of music recommendation, although we expect it to be applicable to most recommendation domains, where item sets are natural units of consumption (e.g., compiling reading lists or healthy dinner recipes).
We implement a high-fidelity multi-modal conversational user interface, where users can listen to songs and leave both item- and set-level feedback.
Using this methodology, we build and publicly release the Conversational Playlist Creation Dataset (CPCD), comprising of over 900 dialogues
containing with 4,800 utterances that express a wide range of preference statements and are paired with over 22,000 item ratings.
See \reffig{example} for an example conversation from the dataset. 

Finally, we use this dataset to experimentally compare a range of conversational retrieval models that are meant to serve as representative baselines. 
Specifically, we focus on the first sub-task of conversational item retrieval and propose an evaluation methodology that adapts standard retrieval metrics to the multi-turn conversational setting.
We find that, while both sparse and dense retrieval methods are useful for this task, our best performing method is a dual encoder architecture~\citep{gtr} pretrained using unsupervised contrastive learning~\citep{Izacard2021-ik} and fine-tuned on the CPCD development set. 
There remains significant headroom for improvement.

In summary, our key contributions are:
\begin{enumerate}
    \item We introduce the task of conversational item set curation, a collaborative task where a user interacts with a system to curate a (small) set of items that meet the user's goal from a (large) collection.
    \item We develop a data collection methodology for conversational item set curation.
    \item We apply this methodology in the music domain to collect the Conversational Playlist Curation Dataset, published at \dataseturl{}.
    \item We present a wide range of conversational retrieval models as baseline approaches and evaluate them on this dataset.
\end{enumerate}

%% file: 02_related.tex
\begin{table*}[ht!]
    \centering
    \begin{tabular}{l c c c c c c}
    \toprule
    \textbf{Dataset}
        & \textbf{\ourdata{}}
        & \textsc{ReDial}~\citep{li2018towards}
        & \textsc{Inspired}~\citep{hayati-etal-2020-inspired}
        & \textsc{DuRecDial}~\citep{liu-etal-2020-towards-conversational,liu-etal-2021-durecdial} 
        & \textsc{GoRecDial}~\citep{kang-etal-2019-recommendation} 
        & \textsc{TaskMaster-2}~\citep{taskmaster2} \\
    \midrule
    Domain(s) &
        Music &
        Movies &
        Movies &
        Movies &
        Movies &
        Music (and others)
    \\
    \# of Dialogs &
    917 & 11,348 & 1,001 & 10,190 & 9,125 & 1,602
    \\
    \# of Turns / Dialog &
    5.8 & 9.1 & 10.7 & 7.6 & 9.9 & 8.9
    \\
    \# of Ratings / Dialog &
    24.5 & 4.45 & 1 & 1 & - & -
    \\
    \# of Candidates &
    106,736 & - & - & 10 & 5 & -
    \\
    Open-ended dialog &
        \cmark{} & \cmark{} & \cmark{} & \xmark{} & \xmark{} & \cmark{} (partial)
    \\ 
    Item set recommendations
    &
        \cmark{} & \xmark{} & \xmark{} & \xmark{} & \xmark{} & \xmark{}
    \\
    Shared item ratings &
        \cmark{} & \xmark{} & \xmark{} & \xmark{} & \xmark{} & \xmark{}
    \\ 
    Item preview &
        \cmark{} & \xmark{} & \cmark{} (post-task) & \xmark{} & \xmark{} & \cmark{}
    \\ 
    Domain-Expert Wizards &
        \cmark{} & \xmark{} & \xmark{} & \xmark{} & \xmark{} & \xmark{}
    \\ 
    \bottomrule
    \end{tabular}
    \caption{
    Comparison of conversational recommendation datasets.
    In contrast to existing datasets,
    \ourdata{} focuses on item set recommendation in the music setting.
    To allow users to provide richer feedback and wizards to respond with more relevant results, we allow them to preview and rate recommendations during the conversation, and employ domain experts as wizards.
    }
    \label{tab:datasets}
\end{table*}

\Section{related}{Related work}

\Subsection{related-datasets}{Conversational Recommendation Datasets}
A key contribution of this work is a new conversational recommendation dataset, the Conversational Playlist Curation Dataset (\ourdata{}).
There are several existing conversational recommendation datasets collected by pairing a ``user'' and ``wizard'' to talk to each other:
    in some settings, as with \ourdata{}, the user has an open-ended dialog with the wizard~\citep{li2018towards,Radlinski2019-ccpe,hayati-etal-2020-inspired,taskmaster2};
    in others users are asked to follow instructions (e.g., mention a specific preference) in each turn of the dialog~\citep{zhou-etal-2020-towards,kang-etal-2019-recommendation,moon-etal-2019-opendialkg,liu-etal-2020-towards-conversational,liu-etal-2021-durecdial,taskmaster2}.
Existing datasets all focus on recommending a single item (typically a movie) to the user;
in contrast, our work focuses on recommending a \textit{set of items} (songs in particular).
To the best of our knowledge, \ourdata{} is the first such dataset.
This distinction leads to several important differences in methodology from prior work:
first, we allow users and wizards to preview recommendations so that they can react accordingly;\footnote{%
While \citet{hayati-etal-2020-inspired} allow users to preview recommended movies at the end of the task prior to rating them, they can't do so during the conversation.
}
second, we allow users to share explicit item ratings with wizards in addition to their responses, leading to more open-ended feedback that often includes soft attributes~\citep{balog2021interpretation}; 
finally, we employ wizards with relevant domain expertise to respond with relevant recommendations.
In addition, \ourdata{} provides a larger set of candidate and relevant items, making it suitable to evaluate the recommendation component of conversational recommendation systems.
\reftab{datasets} summarizes some of the key differences of \ourdata{} with prior work.

An alternative to manually collecting data include synthetic data approaches for conversational recommendation systems (CRSs),
e.g., 
using item rating data~\cite{Dodge2015EvaluatingPQ},
textual review data~\cite{zhang2018towards},
or user logs (e.g., watch history)~\cite{zhou-etal-2020-towards}.
\citet{Dodge2015EvaluatingPQ} and \citet{zhang2018towards} generate dialogs using natural language templates, while \citet{zhou-etal-2020-towards} retrieve candidate utterances and then use human annotators to rewrite them to be more conversational.

\subsection{Conversational Information Seeking}
Beyond recommendation,
conversational information seeking (CIS) systems help users find information through a sequence of interactions that can include search and question answering~\cite{cis-survey}. 
Such systems also seek to enable users to provide direct feedback to improve their results without needing to rely on as much historical interaction data as collaborative filtering-based RSs~\citep{koren2009matrix, koren2022advances}. 
As a result, our work builds on existing work in modeling conversational search and conversational question answering~\cite{yu2021convdr,krasakis-zeroshot,mao-curriculum,zhao2013interactive, zou2020neural, christakopoulou2016towards, lei2020estimation,iai_moviebot}.

\subsection{Entity Retrieval}
Finally, item set recommendation is closely related to entity retrieval~\citep{Balog:2018:EOS}, with 
analogies to entity list completion or example-augmented search~\citep{deVries:2008:OIE,Demartini:2009:OIE,Demartini:2010:OIE,balog11trecentity,balog12trecentity,Zhang:2017:ESA,He:2011:WWW,Bron:2013:EBE,Zhang:2017:SIGIR}.
In entity list completion, the input consists of a textual query and a small set of example entities;
in conversational item set recommendation, the input consists of the conversation history (textual query) and the current item set (example entities).
We follow recent work in dense entity retrieval, and evaluate baselines that use a dual encoder to embed queries and items~\cite{gillick-etal-2019-learning,wu2019zero,leszczynski-etal-2022-tabi}. 

%% file: 03_task.tex
\begin{figure*}[h]
    \centering
    \includegraphics[width=\textwidth]{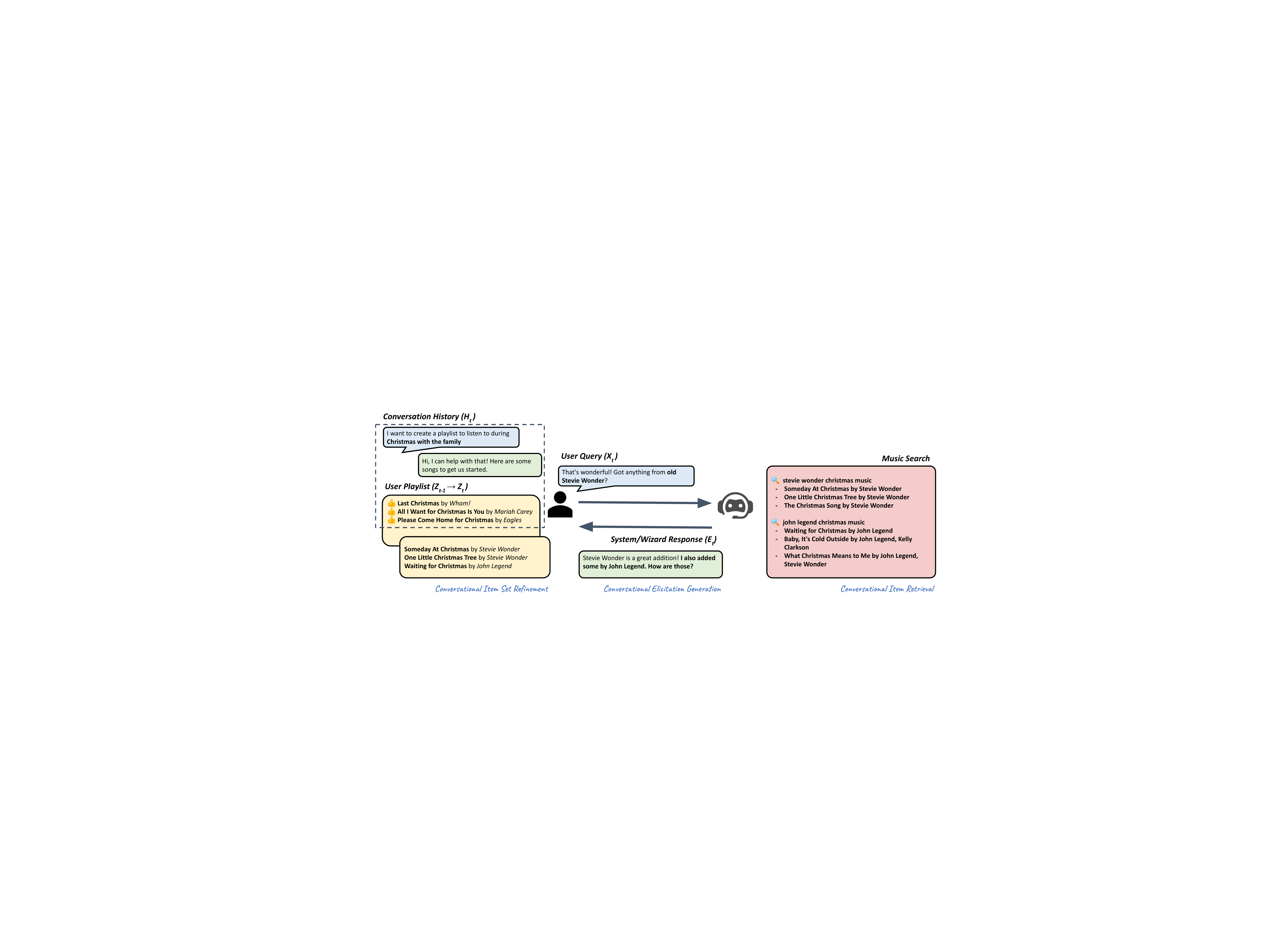}
    \caption{An overview of conversational item set curation (and its subtasks).
    Given a conversation history $H_t$ with previous user utterances $X_{1} \dots X_{t-1}$, item sets $Z_{t-1}$ and system responses $E_{1} \dots E_{t-1}$,
        the system must first retrieve relevant items $Y_t$ (conversational item retrieval),
        then update the item set $Z_t$ (conversational item set refinement)
        and finally generate a response to the user $E_t$ (conversational elicitation / explanation generation).
    }
    \label{fig:cpcd-overview}
\end{figure*}
\Section{task}{Task definition}

Conversational item set curation is a collaborative task where a user interacts with a system to \textit{curate} a \textit{set of items} from a larger collection (\reffig{cpcd-overview}).
We use the term \textit{curate} to emphasize the collaborative and iterative nature of the task:
the goal is to identify not just a single item that matches the user's preferences, but a collection of items that may span a range of interests, but must be ultimately cohesive with the theme or goal of the collection.
While some themes may be narrow (``collection of Elvis Preseley's greatest hits''), others may be broad (``rock music for a long drive'').

In the following, we focus on the music domain where items are songs and the item set is a playlist.
The user begins the task by stating an overall theme or goal for their item set in their first utterance $X_1$, e.g. ``I'd like to create a playlist to listen to during Christmas with the family,''
and the system (possibly proxied by a human wizard) responds with an initial playlist $Z_1$ and an utterance $E_1$ that explains its choice and elicits more information from the user.
In following turns $t$, the user critiques their working playlist $Z_{t-1}$ in their response to the system $X_t$.
The system in turn uses the user feedback to retrieve additional songs $Y_t$ and responds with an updated playlist $Z_t$ and response $E_t$.
Note that the system must decide how many results (if any) from $Y_t$ it should include when updating the playlist $Z_t$.
This process continues for $T$ turns until the user is satisfied with their playlist.

\emph{Conversational item set curation} (CIsC) can be formalized as the task of going from the user goal $X_1$ to an item set that meets their goal $Z_T$.
Given a conversation history, $H_t = (X_1, Z_1, E_1, \dots, X_t)$, a conversational item set curation system must solve three sub-tasks: first, it must retrieve relevant items $Y_t$ from the item corpus (conversational item retrieval), then it must update the item set $Z_t$ (conversational item set refinement) and finally it must generate a response $E_t$ (conversational elicitation / explanation generation).
In this paper, our primary focus will be on conversational item retrieval.

Next, we describe the sub-tasks in further detail:

\paragraph{Conversational Item Retrieval (CIRt)}
In each turn, the system uses the conversation history $H_t$ to retrieve and rank relevant items $Y_t$ from the item collection $\sC$.
To provide contextually relevant recommendations to the user, the system must understand refinements or critiques (``Got anything from \ul{old Stevie Wonder}?'' in \reffig{cpcd-overview}) and soft attributes (``playlist for a sunny day'' in \reffig{example}).
Finally, we note that the scope of CIRt limited to preferences stated by the user thus far; eliciting further preferences is handled below.
We propose evaluating the output of CIRt, $Y_t$, against the target item set $Z_T$ using standard information retrieval metrics such as MRR or Hits@$k$ in \refsec{eval}.

\paragraph{Conversational Item Set Refinement (CIsR)}
Given a list of relevant recommendations from CIRt, the task of CIsR is to identify which and how many of these items should be used when updating the item set.
As reasonable heuristic would add the top $k$ items retrieved by CIRt, but a better method could take into account the confidence and diversity of the retrieved items.
Alternatively, users could be presented all the results from CIRt and be asked to perform item set refinement themselves.
CIsR can be evaluated against $Z_t$ using set-inclusion metrics like precision and recall.
We leave further modeling of this component to future work.

\paragraph{Conversational Elicitation / Explanation Generation (CEEG)}
Finally, user requests are often underspecified;
to provide better recommendations, systems can elicit preferences from users, e.g., by asking them for a genre or probing for why they liked what they like~\citep{Radlinski2019-ccpe}.
In addition, providing an explanation of its recommendation makes the system more scrutable to users and allows them to better steer the system~\citep{scrutable-recommendation}.
While the dataset proposed in this paper includes system responses that exhibit elicitation and explanations, we leave modeling this component to future work.

%% file: 04_protocol.tex
\Section{methodology}{Methodology}

\begin{figure*}
    \centering
    \begin{subfigure}{0.825\columnwidth}
        \includegraphics[height=5.75cm]{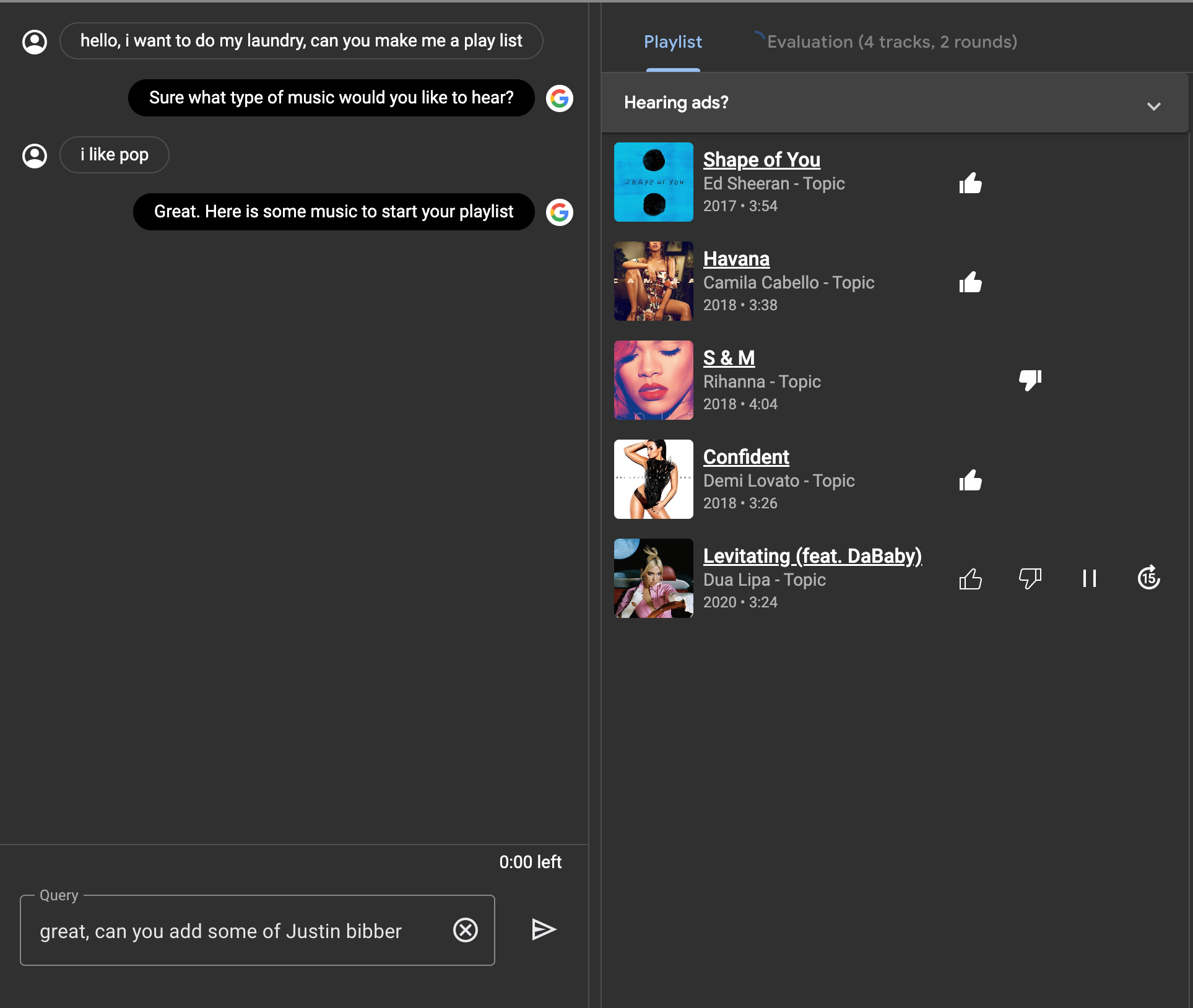}
        \caption{User interface.}
        \label{fig:user-facing-interface}
    \end{subfigure}
    \begin{subfigure}{1.225\columnwidth}
        \includegraphics[height=5.75cm]{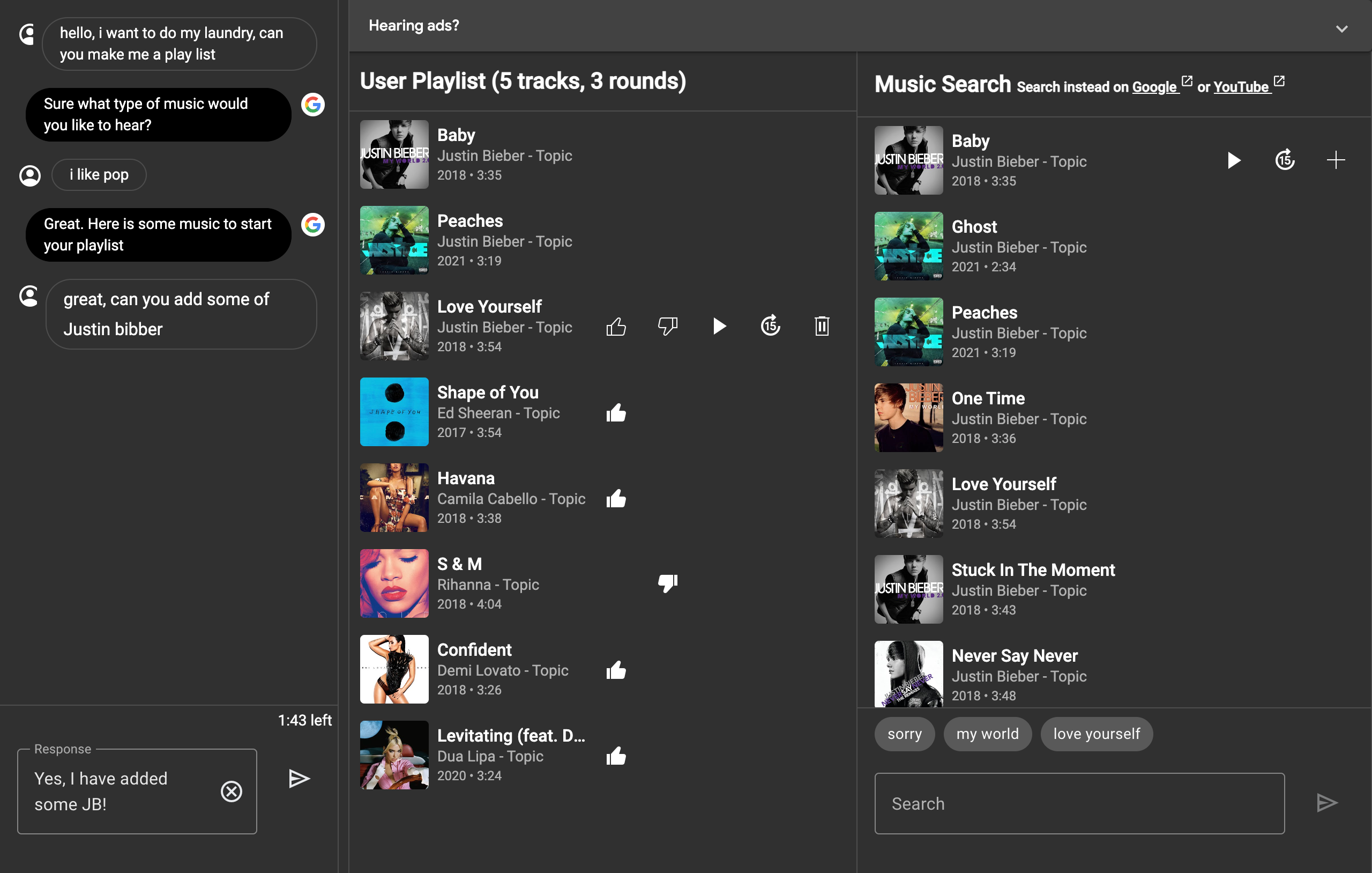}
        \caption{Wizard interface.}
        \label{fig:wizard-facing-interface}
    \end{subfigure}
    \caption{Interface used by users and wizards.
    Users listen to and rate songs in their shared playlist, and respond to the system by communicating how they would like to improve it.
    Wizards are able to see the user's ratings, search for music through the interface and elicit preferences from users.
    }
    \label{fig:interface}
\end{figure*}

Our key contribution is a new dataset, \ourdata{}, to benchmark conversational item set curation systems.
Similar to prior work, we use a human-to-human methodology to collect \ourdata{}: 
    two human raters are recruited to play the role of a user and the system (``a wizard'').
However, to faithfully simulate conversational item set recommendation in consumption domains such as music presents additional challenges:
first, preferences are extremely personal---``pop music'' for one person can be wholly different for another;
second, they are deeply contextual---``pop music'' for a party may be different from ``pop music'' for doing household chores;
third, they are often grounded in the audio content and are hard to understand using only text;
and finally they require substantial domain expertise from wizards to provide relevant recommendations.


\begin{figure*}[]
    \centering
    \begin{subfigure}{\columnwidth}
    \includegraphics[width=\linewidth]{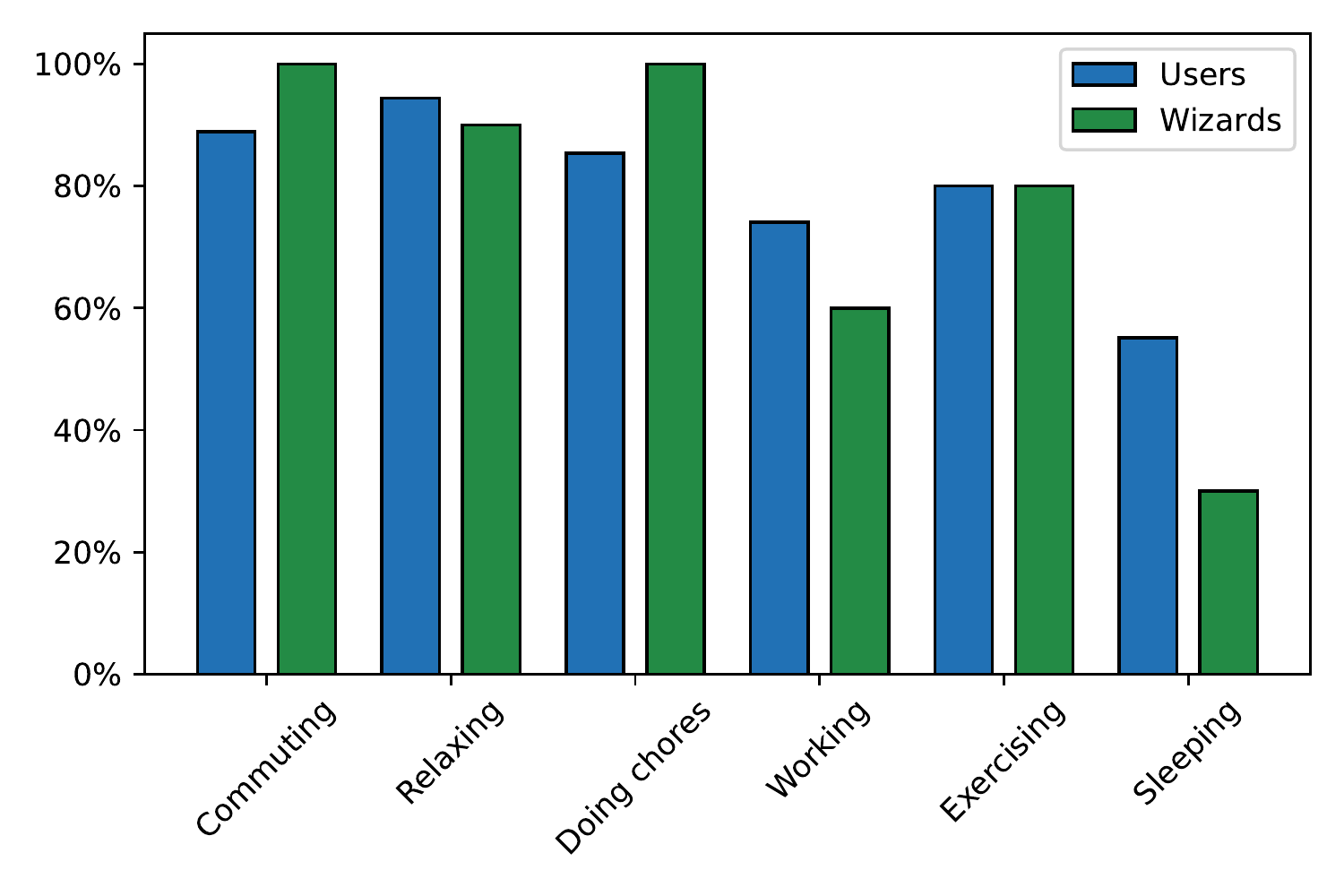}
    \caption{Listening contexts.}
    \label{fig:context-survey}
    \end{subfigure}
    \begin{subfigure}{\columnwidth}
    \includegraphics[width=\linewidth]{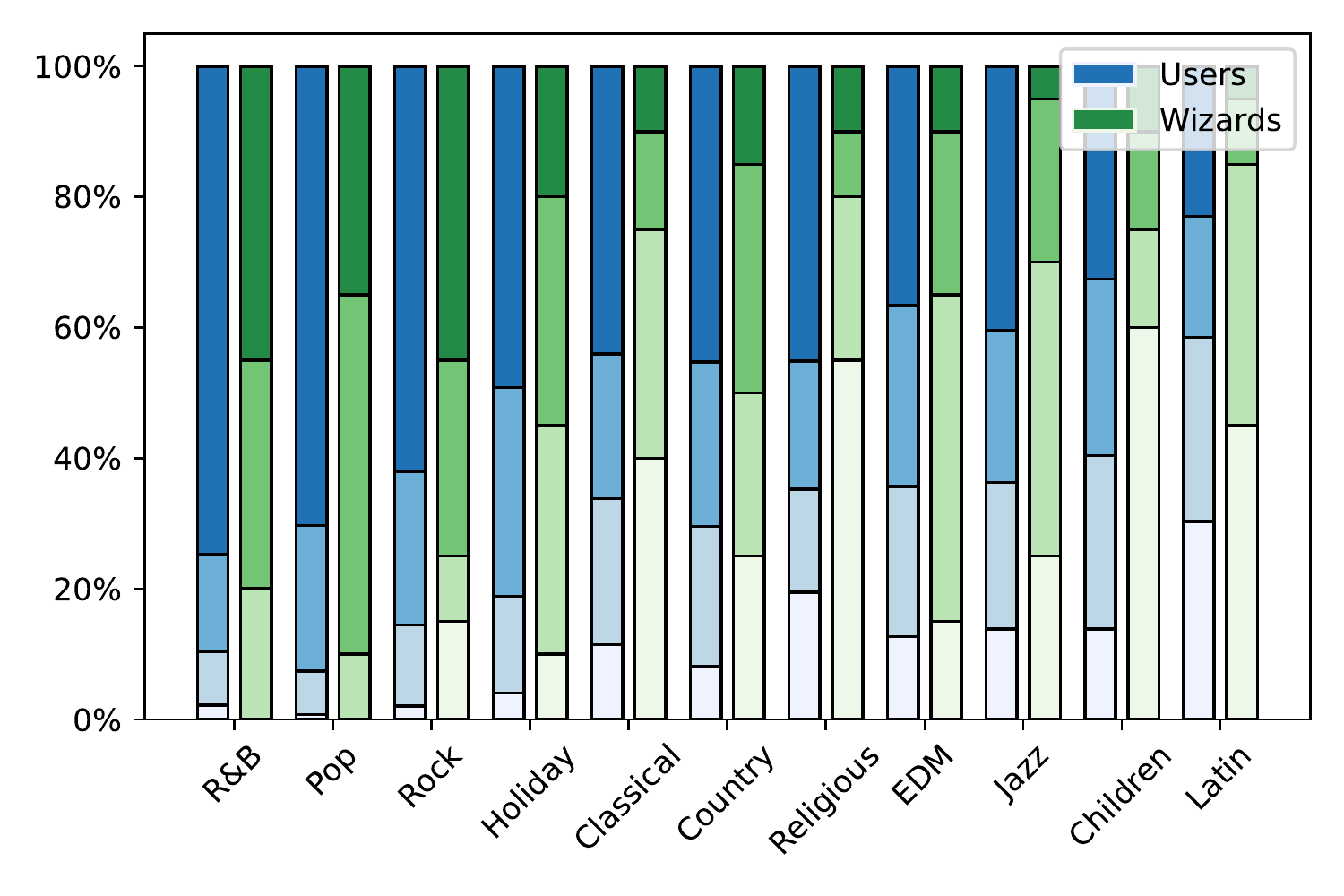}
    \caption{Genre familiarity.}
    \label{fig:genre-survey}
    \end{subfigure}
    \caption{Summary of pre-task music survey results.
    (a) Users and wizards indicated where they listened to music to prime them for the task.
    Most participants listened to music in multiple contexts.
    (b) Users and wizards also indicated how familiar they were with different genres on a 4 point Likert scale (shown with color gradation from "Not at all familiar" being lightest to "Very familiar" being darkest); this information was used to match users and wizards. 
    }
    \label{fig:user-wiz-survey}
\end{figure*}

We address these challenges as follows: we encourage users to personalize the data collection task by asking them to begin each conversation with a specific music-listening scenario in their own lives (e.g., ``walking to work'').
This allows users to relate to the task and fixes the context for the conversation and their listening preferences.
Next, we implement a high-fidelity interface wherein users and wizards can listen to songs during the task.
We found that this was particularly helpful for wizards to better understand users' tastes.
Additionally, we implement a multi-modal interface where users can provide explicit item-level feedback in the form of likes and dislikes, allowing them to focus on broader set-level feedback in their replies.
Finally, in pilot studies, we found that the task required wizards to not only be familiar with music in general, but also with the sub-genres pertinent to the user's interests: wizards with less relevant expertise took longer to perform the task, and had less engaging conversations that were driven entirely by the user.
We address this problem by asking users and wizards to each fill in a survey of their general genre-level preferences, and then matching users with wizards who were most familiar with the users' preferred genres. 

\subsection{Interface design}
A high-fidelity multimodal interface was critical for this complex data collection task. 
The user-facing interface displays the conversational history and the current playlist as recommended by the system (\reffig{user-facing-interface}).
Users can sample songs from their working playlist and provide explicit item-level feedback in the form of likes and dislikes.
Users end their turn by responding to the system via text,
 and are allowed to submit the task after completing at least 5 rounds of conversation and having rated at least 15 songs.
Before submitting the task, users complete a short survey where they rate
    the completeness of their playlist,
    their satisfaction with the playlist,
    the understanding capabilities of the system,
    its helpfulness, and
    their overall experience with the system
    (\reffig{end-task-ratings}).

Wizards are presented a similar interface with an additional element: a music search tool that wizards use to add songs to the user's playlist (\reffig{wizard-facing-interface}).
The music search uses the \href{https://developers.google.com/youtube/v3/docs/search/list}{YouTube Search API}.
We note that the API is limited in its capabilities and tends to return results with a high lexical overlap.
To preserve user privacy, the search results were not personalized.
Wizards were encouraged to search for related artists or recommendations via Google or YouTube Music.

\subsection{Rater Recruitment and Training}
We recruited 10 wizards and 110 users who were fluent English speakers and regularly listened to music from a crowdsourcing vendor.
To control for cultural differences, we required both users and wizards to reside in the U.S.
The wizards were additionally required to have significant music expertise and have performed several music labeling tasks in the past.

Both users and wizards were provided training material in the form of slides.
Users were primed with the following description of the task:
\begin{quote}
    In this study, you will create a \emph{personal} music playlist for a purpose (e.g., “while doing chores” or “when I’m feeling down”) by having a conversation with a system.
\end{quote}
They were then asked to come up with several scenarios for each conversation in preparation for the task.
Users were informed that another person (the wizard) was facilitating the conversation and encouraged to be descriptive in their requests by telling the system what they liked or didn't like, and why.
Users were given instructions on how to use the interface, but were provided no further guidance on what to say.
In contrast, wizards went through multiple pilot rounds of training before beginning the task.
During the pilot rounds, they were paired amongst themselves, and given feedback on how they should respond to users.
During the task, wizards also assisted users with any interface issues (e.g., how to rate items).

Users and wizards were scheduled to interact for a fixed block of time (between 1 and 3 hours), and conducted multiple conversations during this interval.
In order to pair users and wizards, they each completed a short music preferences survey asking them to list where they listen to music (\reffig{context-survey}) and how familiar they are with various genres (\reffig{genre-survey}).
For each user, we ranked wizards based on their overlap with the genres the user expressed familiarity with and tried to pick the top-ranked available wizard for every user subject to scheduling constraints.
Based on pilot feedback, we only matched wizards that were at least somewhat familiar with at least one genre that the user indicated a preference for.
To avoid priming their conversations, users and wizards were not informed of which genres they matched on.
Further personal information about the raters, such as demographics, was not collected as we did not consider it relevant to the task.


\subsection{Post-processing}
After collecting the data, we took several manual and automated steps to process the data.
First, we standardized all white space in the utterances.
Often conversations would begin with the user and wizard checking for each other's presence (e.g., by saying ``hi''); these turns were merged with subsequent ones so that the first turn included the goal statement (e.g., ``Hi, I'd love to create a playlist for classical songs'').
After about 20 minutes, wizards would nudge users to proceed to evaluation:
these utterances were removed along with any other turns at the end of the conversation where no songs were added.
We then manually reviewed the data to identify and remove any turns where users or wizards side-channeled debugging information (e.g., ``I don't see any items,'' ``please rate items so I know what you like'').
Finally, we scraped and included metadata (i.e., titles, artists and album names) for each song mentioned in the dataset.


%% file: 05_analysis.tex
\Section{analysis}{Dataset analysis}

\begin{table}[t]
    \centering
    \begin{tabular}{l r r r}
    \toprule
              & \playlistdata{} & \multicolumn{2}{c}{CPCD} \\
    Statistic & & Dev. & Test \\
    \midrule
    \# of examples & 15,276 & 450 & 467 \\
    \# of tracks in corpus & 332,594 & \multicolumn{2}{c}{106,736} \\
    Avg. \# of turns & - & 5.8 & 5.6 \\
    Avg. $|$query text$|$ & 106.5 & 53.8 & 55 \\
    Avg. $|$response text$|$ & - & 45.6 & 46.0 \\
    Avg. $|$target item sets$|$ & 53.6 & 19 & 18.3 \\
    \bottomrule
    \end{tabular}
    \caption{%
    A summary of key statistics for the Conversational Playlist Creation Dataset (CPCD) and \playlistdata{}, a proprietary collection of expertly curated playlists.
    }
    \label{tab:dataset-stats}
    \vspace{-1em}
\end{table}

\begin{table}[]
    \centering
    \input{figures/preference-examples.table}

    \caption{Examples of preferences in the dataset and their estimated frequency.
    The dataset includes rich preference statements: over 30\% of user preferences include soft attributes such as activities, sounds and moods.
    Many artist preferences use artists as a point of reference (e.g, ``Lonnie Liston is nice. More similar artists would be good.'').
    }
    \label{tab:preference-examples}
\end{table}

\begin{table}[]
    \centering
    \input{figures/examples.tex}
    \caption{
    Examples of conversations in the dataset.
    The examples showcase the breadth of music interests and listening contexts in the dataset, and how users center thematic cohesiveness and diversity in their preferences.
    Due to space considerations, only the first three turns and up to three results from each turn are shown.
    }
    \label{tab:examples}
\end{table}

We now take a deeper look at the \ourdata{} dataset.
In total, after filtering and post-processing, the dataset contains 917 conversations that are randomly split into development and test sets; see~\reftab{dataset-stats}.
\reftab{examples} includes a few example conversations from the dataset. 
On average, each conversation includes about 5.7 turns each for the user and wizard, and ends with a final playlist of about 19 songs.
Users rate about 4.8 songs in each turn, of which about 20\% are negative ratings.
In total, the dataset includes over 100k unique songs that serve as the item corpus for retrieval tasks.

\begin{figure*}
    \centering
    \begin{subfigure}{\columnwidth}
    \includegraphics[width=\linewidth]{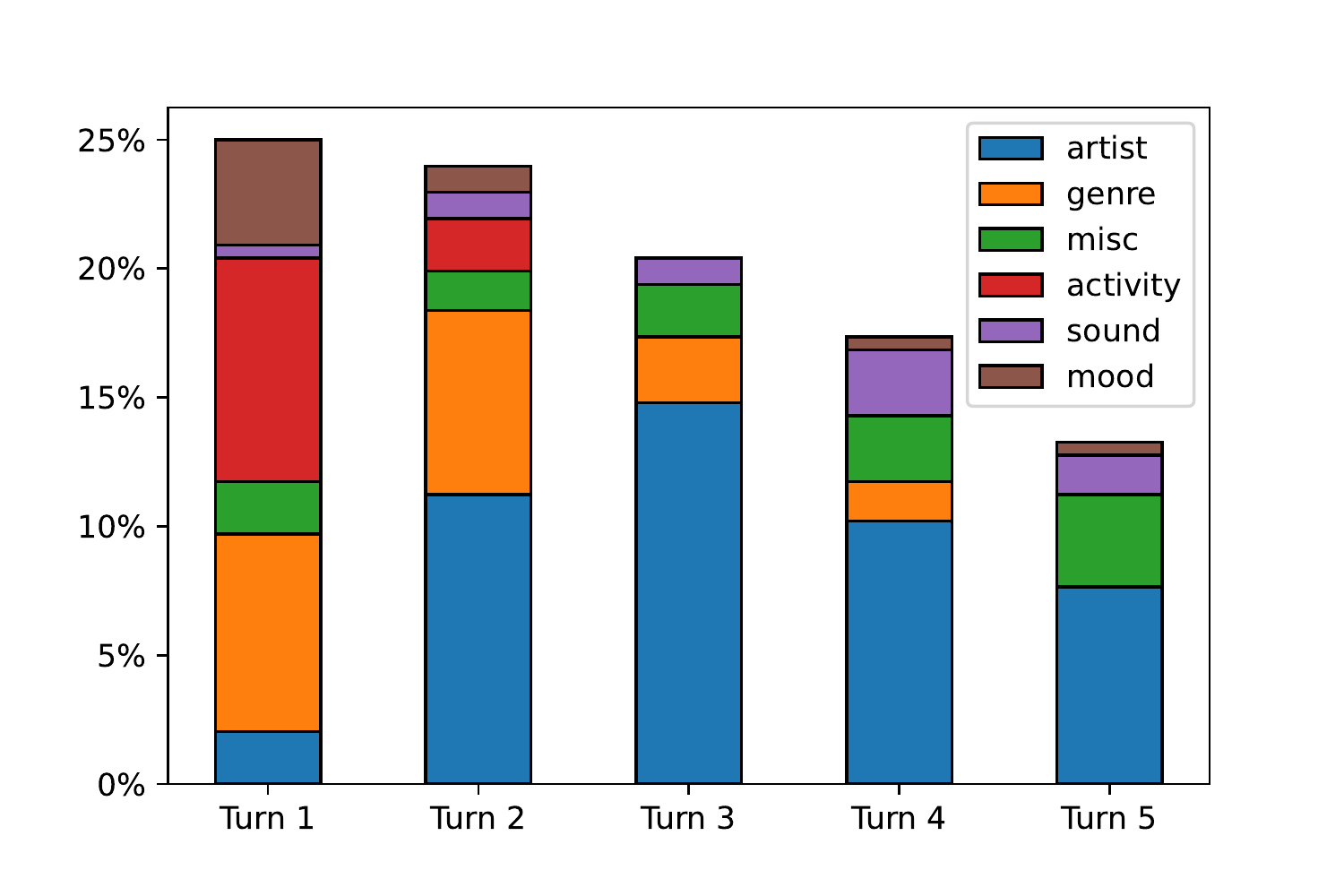}
    \caption{User Preferences.}
    \end{subfigure}
    \begin{subfigure}{\columnwidth}
    \includegraphics[width=\linewidth]{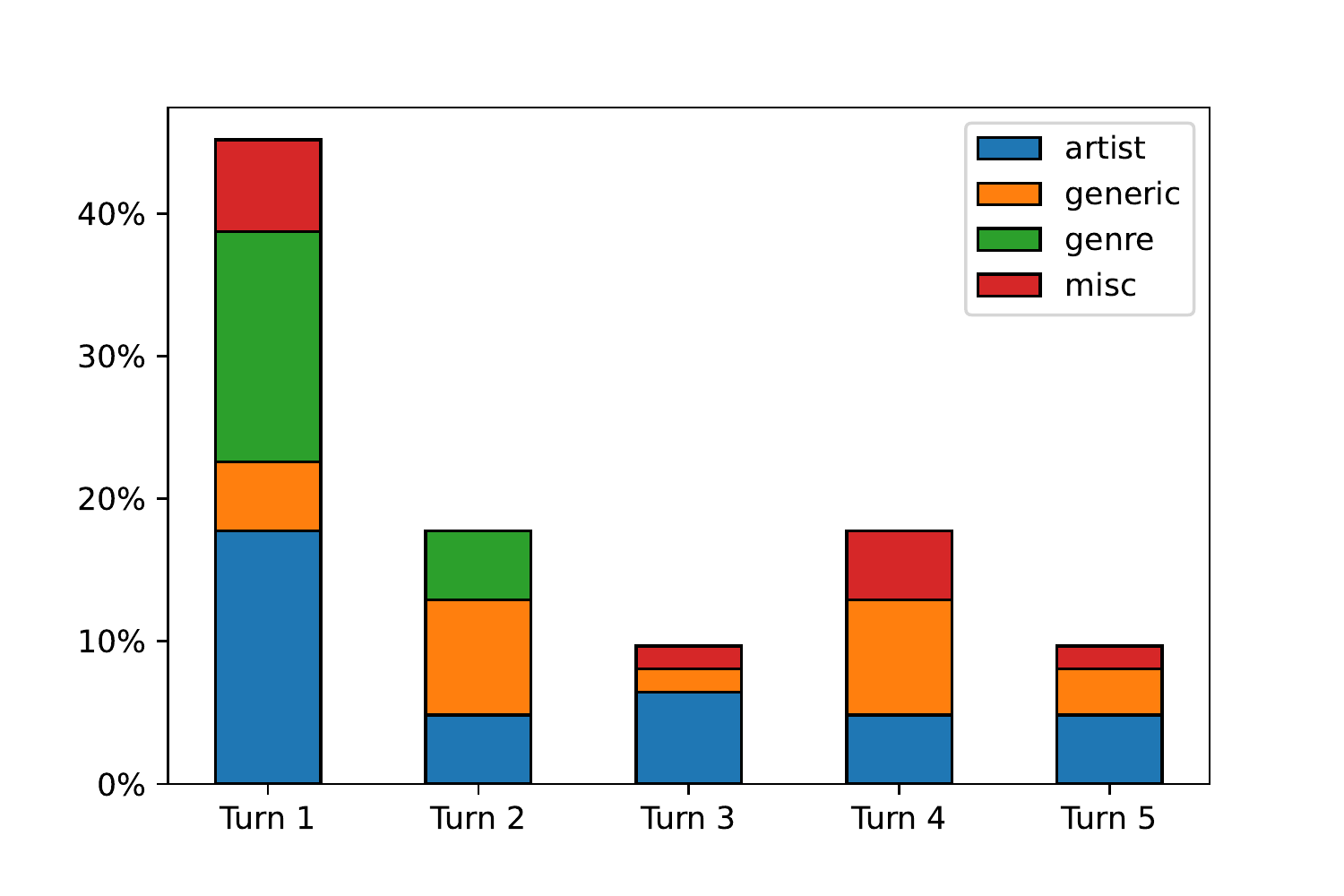}
    \caption{Wizard Preferences.}
    \end{subfigure}
    \begin{subfigure}{\columnwidth}
    \includegraphics[width=\linewidth]{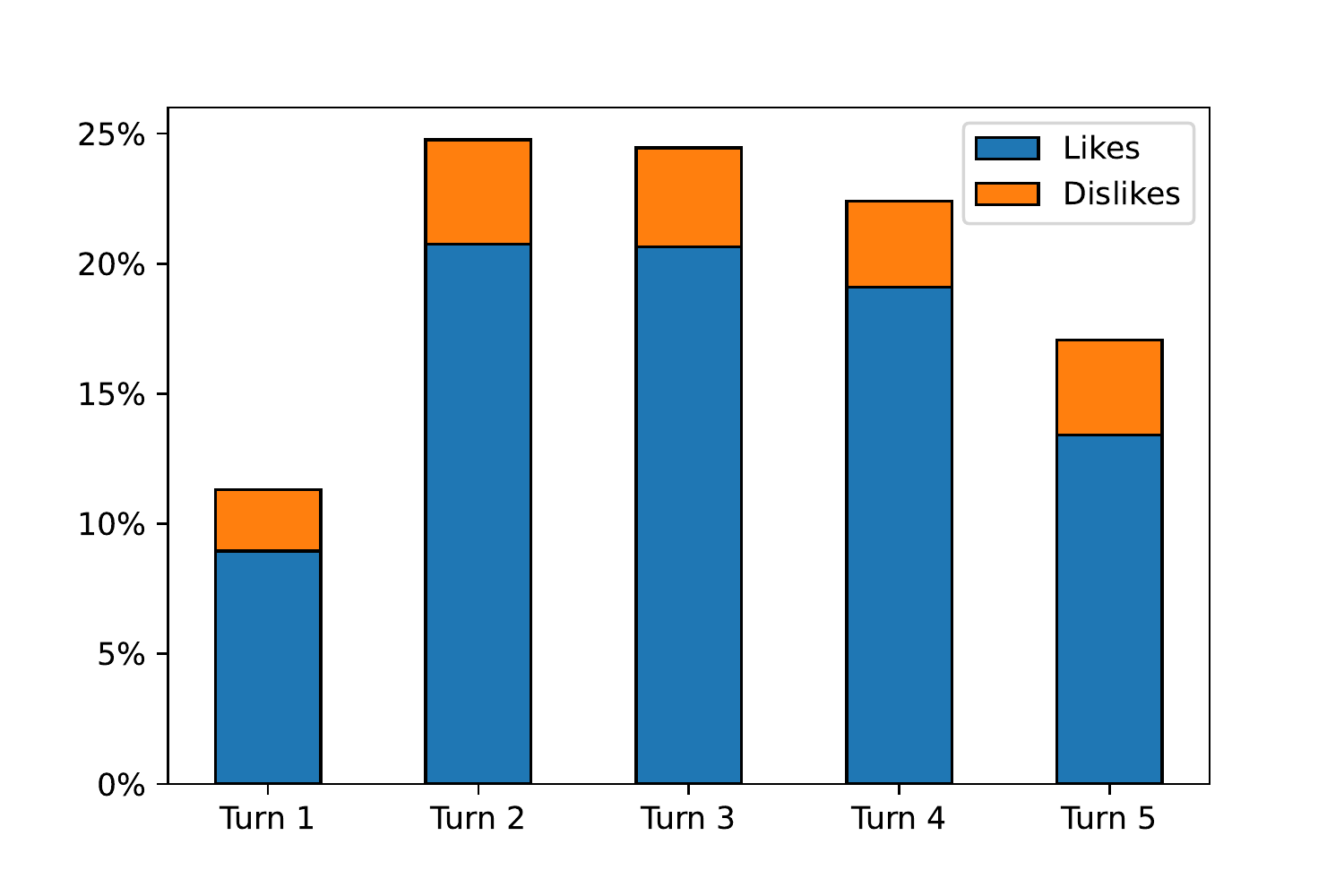}
    \caption{Per-turn Item Ratings.}
    \label{fig:per-turn-ratings}
    \end{subfigure}
    \begin{subfigure}{\columnwidth}
    \includegraphics[width=\linewidth]{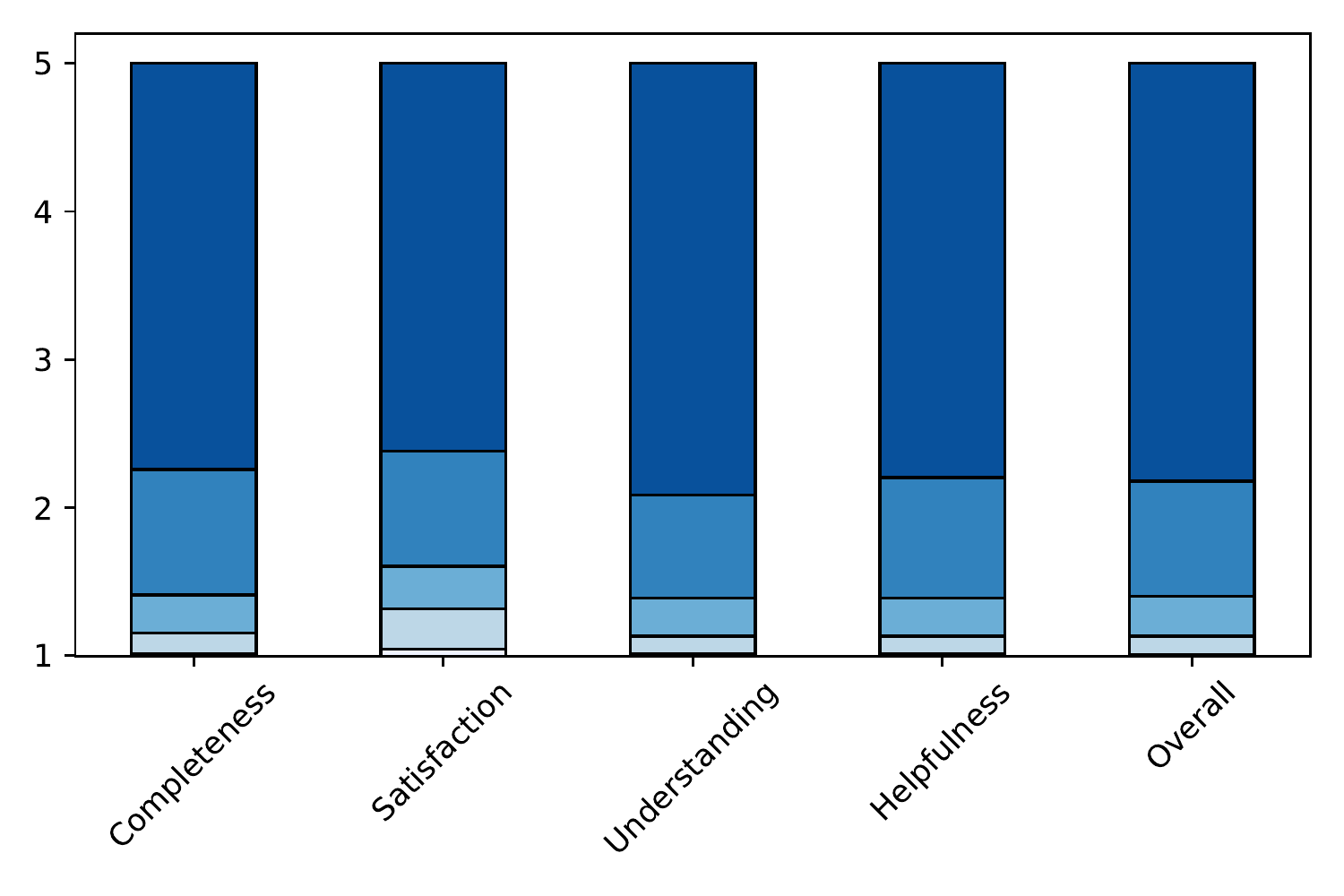}
    \caption{Post-task User Ratings.}
    \label{fig:end-task-ratings}
    \end{subfigure}
    \caption{
    (a, b) Preferences expressed by users and elicited by wizards through their conversations. Users often start the conversation with activity and genre preferences, which are maintained as part of context, and then provide additional artist based preferences (often as examples) in later turns.
    Likewise, wizards begin conversations by eliciting users for artist and genre preferences.
    }
    \label{fig:preferences-by-turn}
\end{figure*}

Next, we analyze the distribution of preferences expressed by users and elicited by wizards. 
To do so, we manually annotated 220 turns from 41 conversations.
After exploratory analysis, we identified 6 key categories of preferences for users and and 4 for wizards;  \reftab{preference-examples} lists these categories with typical examples.
While artist and genre based preferences make up a majority of the dataset, over 30\% of preferences include soft attributes such as activities, sounds, moods, and more.
Additionally, many artist preferences use artists as a point of reference (e.g., ``Maybe something from 2pac and Ice Cube?'').
Users express preferences in 80\% of turns, while wizards elicit preferences 29\% of the time.
We also observed that only about 5\% of preferences in the dataset are negative preferences, e.g. ``Those might be too similar, though I do prefer them to \ul{not have any lyrics}.''

\reffig{preferences-by-turn} shows how the relative frequency of different preferences change across turns.\footnote{%
We only show the first 5 turns in \reffig{preferences-by-turn} as the number of data points sharply drops after 5 turns.
}
Activity and genre preferences are often expressed early in the conversation, and maintained as part of the conversation context in later turns.
Wizards often begin by eliciting the user for their preferences, and thus add fewer songs to the playlist in the first turn (\reffig{per-turn-ratings}).
We note that wizards often elicited users for artists and genres, which were the attributes that worked most reliably with the search tool; we believe this played a role in the wizards' choices.

Finally, \reffig{end-task-ratings} summarizes the distribution of post-task user ratings.
Users were overall satisfied by the experience and 
    expressed that the system understood their preferences well and was very helpful in assisting them on the task.
They also expressed that they thought their playlist was mostly complete, and that they would listen to it again.
The post-task ratings were weakly correlated with the match scores, with Spearman rank correlation ranging from 0.11 for playlist satisfaction to 0.19 for system helpfulness ($p < 0.05$ in all cases), suggesting marginal benefit beyond ensuring wizards were at least somewhat familiar with genres the user was familiar with.

%% file: figures/preference-examples.table.tex
    \begin{tabular}{l p{5.7cm}}
    \toprule
    Type (Freq.\%) & Example \\
    \midrule
    \multicolumn{2}{l}{\textit{Preferences expressed by users}} \\
        Artist (46.0\%) & My favorite artist is \ul{Drake}. \\
        Genre (18.9\%) & I love \ul{old school rap}... \\
        Activity (10.7\%) & I'd like to create a playlist that I can utilize \ul{for a dance party}. \\
        Sound (6.6\%) & Anything with a \ul{long melodic hook} is not good for my workouts. \\
        Mood (6.1\%) & So Ambitious is too slow paced for workouts as well. Specifically the hook is \ul{not upbeat enough}. \\
        Others (11.7\%) & More gospel music \ul{from the late 2000s}. \\
                        & I'd love some songs by \ul{British artistes}. \\
        \\
    \midrule
    \multicolumn{2}{l}{\textit{Preferences elicited by wizards}} \\
        Artist (38.7\%) & Of course, who are your favorite artists? \\
                        & You got it. How about some J Cole or Kendrick? \\
        Results (25.8\%) & Are any of these good? \\
                        & What about these ones? \\
        Genre (21.0\%) & Do you like soft rock, Jazz, pop, classical, or country sounds? \\
        Others (14.5\%) & Are you trying to focus as you reorganize? \\
                        & Any particular era or country of origin\dots? \\
                        & Do you prefer the slower ones? \\
    \bottomrule
    \end{tabular}

%% file: figures/examples.tex
\footnotesize{}
\begin{tabularx}{\columnwidth}{l p{0.78\columnwidth}}
\toprule
User & I want to create a playlist for a girls night with some music from Nicki Minaj. \\
System & Got you! What do you think about these songs by Nicki? \\
    & \cmark{} Feeling Myself by Nicki Minaj, Beyoncé \\
    & \cmark{} Beez In The Trap by Nicki Minaj, 2 Chainz \\
    & \xmark{} Super Bass by Nicki Minaj \\
User &  Yes that is exactly what I am looking for \\
System & Perfect! What about these? \\
    & \cmark{} Bartier Cardi by Cardi B, 21 Savage \\
    & \xmark{} Ain't Sh*t by Doja Cat \\
    & \xmark{} Big Energy by Latto \\
User & I definitely would like to have more Cardi B on this playlist.
    Not liking the songs by the other artists too much. \\
System & No worries, here are some songs by Cardi B \\
    & \cmark{} WAP by Cardi B, Megan Thee Stallion \\
    & \cmark{} Thru Your Phone by Cardi B \\
    & \cmark{} Money by Cardi B \\
\midrule{}
User & I want to create a playlist for working out \\
System & Of course, who are your favorite artists? \\
User & I enjoy working out to Hip Hop music. Some of my favorite artists are JayZ and Eminem \\
System & You got it \\
& \cmark{} The Way I Am by Eminem \\
& \cmark{} So Ambitious by Jay-Z, Pharrell Williams \\
& \xmark{} Young Forever by Jay-Z, Mr Hudson \\
User & The song "Young Forever" is too slow paced for a workout playlist. Great song though. \\
System & Here's a few uptempo Jay-Z songs for you \\
& \cmark{} On To The Next One by Jay-Z, Swizz Beatz \\
& \cmark{} I Got the Keys by DJ Khaled, Jay-Z, Future \\
& \xmark{} Hey Papi by JAY-Z, Memphis Bleek, Amil \\
User & So Ambitious is too slow paced for workouts as well. Specifically the hook is not upbeat enough. \\
System & Here's a few more options for you \\
& \cmark{} Run This Town by Jay-Z, Kanye West, Rihanna \\
& \cmark{} GOD DID by DJ Khaled, Rick Ross, Lil Wayne, Jay-Z, John Legend, Fridayy \\
& \cmark{} D.O.A. (Death Of Auto-Tune) by Jay-Z \\
\midrule{}

User & I'd like to create a playlist that will put me to sleep, maybe some instrumentals \\
System & Sounds great! Here's a few to start \\
& \cmark{} Oase by Mo Anando \\
& \cmark{} Rainbow Roll by Strehlow \\
& \xmark{} throwback port by mommy, Philanthrope \\
User & those are pretty songs \\
System & Here are a few more along a similar vein for your review \\
& \cmark{} Just Chilling by Grisp \\
& \cmark{} Life Circle by mr. käfer \\
& \xmark{} Raindrops Keep Fallin' by Phlocalyst \\
User & Those might be too similar, though I do prefer them to not have any lyrics \\
System & Ok, here are some very different instrumentals \\
& \cmark{} Black Mountain Side (Remaster) by Led Zeppelin \\
& \xmark{} Electric Worm by Beastie Boys \\
& \xmark{} Frankenstein by The Edgar Winter Group \\
\bottomrule
\end{tabularx}

%% file: 06_baselines.tex
\Section{baselines}{Modeling Conversational Item Retrieval}
We focus on modeling conversational item retrieval, and leave conversational item set refinement and conversational elicitation and explanation generation to future work.
In particular, we consider three canonical retrieval modeling approaches: sparse retrieval, dense retrieval, and query rewriting.

\Subsection{sparse-model}{Sparse Retrieval: BM25}
BM25 is a commonly used sparse, bag-of-words ranking function that measures the relevance of a document or $x$ to a query $q$.
We use BM25 in a sparse retrieval system by treating each item in the retrieval corpus as a candidate document represented in text as ``\texttt{<title> by <artist> from <album>},'' and returning the top ranked items for each query.\footnote{%
We used the hyperparameters $k_1 = 1.5$ and $b=0.75$ in our implementation.
}
We include conversation history in queries by concatenating user queries from previous turns (omitting conversation history led to worse performance).
For consistency with the dense retrieval methods below, we use SentencePiece~\citep{kudo2018sentencepiece} to tokenize both queries and documents.

\Subsection{model-inputs}{Dense Retrieval Methods}
We consider a popular and efficient dense retrieval method: a dual encoder~\citep{gillick-etal-2019-learning,dpr,gtr}. 
Dual encoders independently embed queries $q$ and items $x$ into normalized dense vectors using a neural network,
and retrieve the most similar items for a query using cosine similarity:
$\rho(x; q) = f(q)^\top g(x)$,
where $f: \sQ \to \R^d$ and $g: \sX \to \R^d$  are embedding functions for queries and items, respectively.

At turn $t$, we represent the using the conversation history $H_t = (X_1, Z_1, E_1, \dots, X_t)$ in text by concatenating user utterances with text representations of the top-k ($k=3$) songs added to the item set in each turn in reverse chronological order.
For example, the query at turn $t$ is: 
\begin{equation*}
X_t \sep d(\delta Z_{t-1}) \sep X_{t-1} \dots~d(\delta Z_1) \sep X_1,
\end{equation*}
where $d(\delta Z_t)$ is a textual representation of the items (same as in BM25) added to the item set in turn $t$, and \sep represents a separator token.
We do not include the previous system response ($E_t$) in the query.

\subsubsection{Training and Inference}


For training, we use a standard contrastive loss with in-batch negatives:
given an example $i$ with query $q_i$ and target item set $Z_i$,
we randomly sample an item from $x_i \in Z_i$ to be a positive target,
and take items from other item sets in the batch, $x_j$ where $j \neq i$, to be negative targets.
Additionally, we augment the training data to improve robustness as follows: 
(1) we generate conversations of varying lengths by randomly truncating conversations to its first $t$ turns, and
(2) we generate historical item sets of varying lengths by randomly truncating them to their first $k$ items.

\subsubsection{Model implementation.}
\label{sec:model-details}
We use a shared encoder for $f$ and $g$ initialized either from a pre-trained T5 1.1-Base checkpoint~\citep{JMLR:v21:20-074} or a T5 1.1-Base checkpoint that has been further pre-trained on retrieval tasks using the Contriever objective~\citep{Izacard2021-ik}.
We then fine-tune the encoder on the \ourdata{} development set for 1,000 steps using Adafactor~\cite{Shazeer2018-pm} with a batch size of 512 and a constant learning rate of 1e-3 (further fine-tuning led to overfitting).
We denote these methods as DE$\rhd$\ourdata{} and Contriever$\rhd$\ourdata{} respectively.

Additionally, to explore other relevant training data, we also fine-tuned a T5 1.1-Base checkpoint on a proprietary collection of 15,276 expertly curated playlists, \playlistdata{}, by using its playlist descriptions as queries and the playlists themselves as the target item sets.
We denote this system as \nonconvo{}.

For inference, we build an index of pre-computed item embeddings using each method's respective encoder.
We embed queries as in training, and use nearest neighbor search to return the top-$k$ items for $q_t$.

\Subsection{query-rewriting}{Query rewriting methods}
Query rewriting or de-contextualization rewrites a query to incorporate conversation history, and has been widely used in state-of-the-art conversational search systems~\cite{vakulenko-question-rewriting,yu-fewshot,Lin2020-cq}.
Following prior work~\citep{Lin2020-cq}, we fine-tune a T5-based query rewriter on the CANARD dataset~\cite{elgohary2019canard} and use it to rewrite the queries in CPCD.
We adapt two of the above methods, BM25 and \nonconvo{}, to use a query rewriter by replacing the conversation history with rewritten queries; the resulting methods are named BM25+QR and \nonconvo{}+QR respectively.

\Section{eval}{Evaluating Recommendation Performance}
We now use \ourdata{} to evaluate the conversational item retrieval methods described in \refsec{baselines}.

\Subsection{exp-setup}{Experimental Setup}
\subsubsection{Baselines.}
We evaluate the following baseline methods: BM25, BM25+QR, \nonconvo{}, \nonconvo{}+QR, DE$\rhd$\ourdata{}, and Contriever$\rhd$\ourdata{}.
Additionally, we include a popularity baseline that returns the top 100 songs from the development set for every query in the test set.

\subsubsection{Evaluation metrics.} 
\label{sec:eval-metrics}
Evaluating recommendation systems is challenging because 
there are often missing ratings~\cite{Dodge2015EvaluatingPQ,ValcarceMetrics2018}: 
we do not know whether an unrated song would be liked or disliked by a user. 
Standard ranking-based metrics treat items without ratings as not relevant---even though they may actually be relevant---and thus provide a lower bound on the recommendation performance.
Following prior work~\cite{Dodge2015EvaluatingPQ,zhou-etal-2021-crslab,kang-etal-2019-recommendation}, we compare systems using a standard ranking metric, Hits@$k$,
which is 1 if and only if any of the top-k retrieved songs are in the target item set.
Unless otherwise stated, we report a macro-averaged Hits@$k$, by averaging Hits@$k$ across turns within a conversation and then across conversations.
We report Hits@$k$ for multiple values of $k$ (10, 20, 100):
    while smaller values are more representative of realistic scenarios where users only consider a small set of items, \citet{ValcarceMetrics2018} found that larger values of $k$ have more discriminative power in offline experiments and can help mitigate the missing rating problem.

Evaluating \textit{conversational} recommendation systems poses an additional challenge: how to support evaluation across multiple turns.
Different systems may recommend different items in previous turns, leading to divergent histories.
To fairly compare across models within a turn, we assume a shared ``gold'' history and target item set across all models. 
Additionally, we remove any items in the target item set that also appear in the gold history.
When there are additional songs liked in a turn, but not added to the history (e.g., when there is a limit to the number of songs in the history), we keep them in the target item set as leftovers~\cite{paraparleftovers}, resulting in a larger target item set.

\subsection{Main Results} 
\begin{table}[t]
\input{figures/benchmark_results.table}
\caption{
Hits@$k$ on CPCD test. DE denotes dual encoder; QR denotes query rewriter.
Underlined numbers denote statistical significance compared to the popularity baseline, according to a paired randomization test ($p < 0.05$).
}
\label{tab:benchmark_results}
\vspace{-1em}
\end{table}

\reftab{benchmark_results} compares models on the CPCD test set.
As expected, the popularity baseline is among the worst evaluated, and its performance reflects the degree of overlap between the development and test sets: about 26\% of songs liked in the development set were also liked in the test set and 38\% of the top 100 songs in the development set also appear in top 100 songs of the test set.
Next, the sparse retrieval baseline, BM25, does significantly better than the popularity baseline and is the second best method overall. 
In particular, we find that BM25 scores highly on artist-based queries which are well-suited to sparse bag-of-words based methods. 
Furthermore, the search tool used by wizards appears to rely on lexical overlap, which may bias the results in favor of sparse methods.
We found that using query rewriting either did not improve or hurt performance: we attribute this to the rewriter not generalizing well to the music-seeking queries in CPCD; better query rewriting methods are a promising avenue for future work.
Finally, the performance of the dense retrieval methods depended on both the fine-tuning dataset as well as the pretraining strategy:
    while training on playlist descriptions (DE$\rhd$\playlistdata{}) outperforms the popularity baseline, it still lags far behind BM25,
    and training solely on the \ourdata{} (DE$\rhd$\ourdata{}) does even worse due to insufficient training data.
In contrast, training on \ourdata{} from a model pretrained for the retrieval task (Contriever$\rhd$\ourdata{}) is the best method we evaluated and significantly outperforms BM25 ($p < 0.05$).

\paragraph{Recommendations on using CPCD for model training.}
Given its relatively small size, it can be challenging to train a dense retrieval model using CPCD alone (e.g., DE$\rhd$\ourdata{}), and recommend using models pretrained on other resources (e.g., Contriever$\rhd$\ourdata{}).
When using the dataset for training, we provide canonical train and validation splits of the development set, but encourage e.g., $k$-fold cross-validation to use the data more efficiently.

%% file: figures/benchmark_results.table.tex
\begin{tabular}{lrrr}
\toprule
Model                                     & Hits@10  & Hits@20  & Hits@100 \\
\midrule
Popularity                &  7.8 & 14.8 & 37.9 \\
BM25                      & \ul{19.7} & \ul{27.4} & \ul{45.5} \\
BM25+QR                   & \ul{15.5} & \ul{21.7} & \ul{34.0} \\
DE$\rhd$\playlistdata{}
                          & \ul{13.1} & \ul{19.6} & \ul{43.2} \\
DE$\rhd$\playlistdata{}+QR
                          & \ul{13.1} & \ul{20.2} & \ul{42.5} \\
DE$\rhd$\ourdata{}
                          & \ul{10.7} & \ul{16.1} & \ul{30.1} \\
\textbf{Contriever$\rhd$\ourdata{}}
                          & \ul{\bf{27.5}} & \ul{\bf{36.2}} & \ul{\bf{49.6}} \\
\bottomrule
\end{tabular}


%% file: 07_discussion.tex
\Section{conclusion}{Conclusion}

We introduced the task of conversational item set curation to capture recommendation scenarios, such as music consumption, where users are looking for a set of items instead of individual ones.
In these settings, we imagine users to collaborate with a system to iterative curate an item set through conversation.
Given a conversation history, we subdivided the task into conversational item retrieval (CIRt), conversational item set refinement (CISR) and conversational elicitation and explanation generation (CEEG). 
To facilitate research in conversational item set curation and its subtasks, we developed an efficient data collection methodology and used it to create the Conversational Playlist Curation Dataset (CPCD).
The dataset contains rich preference statements from both users and wizards that are unique to the item set recommendation setting.
Finally, we used the data to evaluate a wide range of conversational item retrieval methods.
Specifically, our evaluation focused on the CIRt subtask, where we showed that both sparse and dense retrieval methods are useful.
Further modeling---e.g., pretraining paradigms that target conversational music recommendation or  explicitly modeling audio content---is necessary to improve performance on this task.
In future work, we plan to explore using CPCD how to model and evaluate the remaining subtasks, CISR and CEEG, by utilizing the search actions and responses collected from wizards.

